\documentstyle[prl,aps,epsfig,prbbib]{revtex}
\begin{document}
\draft
\twocolumn[\hsize\textwidth\columnwidth\hsize\csname  
@twocolumnfalse\endcsname

\title{Time-dependent Gross-Pitaevskii equation for composite bosons
as the strong-coupling limit of the fermionic BCS-RPA 
approximation}
\author{G.C. Strinati and P. Pieri}
\address{Dipartimento di Fisica, UdR INFM,
 Universit\`{a} di Camerino, I-62032 Camerino, Italy}
\date{\today}
\maketitle
\hspace*{-0.25ex}

\begin{abstract}
The linear response to a space- and time-dependent external disturbance of
a system of dilute condensed composite
bosons at zero temperature, as obtained from the linearized version of the
time-dependent Gross-Pitaevskii
equation, is shown to result also from the strong-coupling limit of the
time-dependent BCS (or broken-symmetry RPA)
approximation for the constituent fermions subject to the same external
disturbance. In this way, it is possible to connect excited-state properties
of the bosonic and fermionic systems by placing the Gross-Pitaevskii 
equation in perspective with the corresponding fermionic approximations.
\end{abstract}
\hspace*{-0.25ex}
]
\narrowtext
Ultracold atomic Fermi gases are currently being intensively studied 
experimentally, as their low-temperature
properties can shed light on fundamental questions regarding degenerate
Fermi systems.
Producing a paired superfluid is a particular challenge, and is being  
pursued by several groups \cite{DM-Jin-99,Inguscio-01,O-Hara-02}.
Since the attractive interaction responsible for superfluidity
in a Fermi gas can be tuned to reach the strong-coupling condition, it also 
appears possible to study
experimentally the evolution
from superfluid fermions to condensed composite bosons. (These form as
 bound-fermion pairs in the strong-coupling limit of the fermionic 
attraction.) This would permit for the first time a controlled 
\emph{crossover\/} 
between two basic quantum systems sharing the same spontaneous
broken-symmetry (superfluid) behavior.

Dilute condensed bosons have already been studied extensively both
experimentally \cite{dilute-bosons} and
theoretically \cite{DGPS-99-Leggett}.
The diluteness condition is matched in almost all
current experiments. Correspondingly,
the Gross-Pitaevskii (GP) equations \cite{Gross,Pitaevskii} have proven
sufficient to describe the observed phenomena. This is true
both for ground-state properties (via the time-independent GP equation)
\emph{and\/} for the dynamical response of the
condensate to an external disturbance (via the time-dependent GP equation in
its linearized form \cite{Pethick-Smith}).

With the aim of connecting properties of superfluid fermions and condensed
composite bosons, the time-independent GP equation for the condensate wave
function was obtained in a previous paper
\cite{PS-02} as the strong-coupling
limit of the Bogoliubov-de Gennes equations \cite{DeGennes} for superfluid
fermions at zero temperature.

In this paper, a corresponding connection is established between the
linearized form of the time-dependent GP equation
(from which the dynamics of the bosonic condensate can be obtained) and the
strong-coupling limit of the linear response
for superfluid fermions with a BCS ground state, subject to the same
external disturbance.
We will explicitly show that the fermionic density and current correlation
functions map, in the strong-coupling limit of the
fermionic attraction, onto the linear-response results for condensed 
composite bosons. The fermionic correlation functions are obtained from the 
time-dependent BCS (or broken-symmetry RPA) approximation, while the 
linear-response results for the composite bosons are obtained from the 
time-dependent GP equation.
Thus a connection between ground- and excited-state properties of
the condensate both in the weak- and
strong-coupling limits is established.
On physical grounds the connection between the time-independent and
time-dependent versions of the GP equation for condensed composite bosons and 
the BCS and time-dependent BCS approximations for superfluid
fermions, rests on the zero-point motion in a quantum
theory implying an oscillatory spectrum.
The imprint of the quasi-particle spectrum should be
found in the ground-state wave function\cite{Gross}.

The response functions of a system of condensed bosons described by the
wave function
$\Phi({\mathbf r},t)$ can be obtained from the time-dependent GP equation
\cite{Gross,Pitaevskii}:
\begin{eqnarray}
&&\frac{1}{2m_{B}}  \left( i \nabla + \frac{q_{B}}{c} 
{\mathbf A}_{{\rm ext}}({\mathbf r},t) \right)^{2}\Phi({\mathbf r},t)
 +  V_{{\rm ext}}({\mathbf r},t)\Phi({\mathbf r},t)\nonumber\\ 
&&\;\;\;\; +\, U_{0} |\Phi({\mathbf r},t)|^{2} \Phi({\mathbf r},t)
=  i \, \frac{\partial \Phi({\mathbf r},t)}{\partial t}
\label{time-dependent-GP}
\end{eqnarray}
in the presence of space- and time-dependent vector
${\mathbf A}_{{\rm ext}}$ and scalar $V_{{\rm ext}}$ external
potentials.
Here, $c$ is the velocity of light, $m_{B}$ and $q_{B}$
are the mass and charge of the
bosons, and $U_{0} = 4 \pi a_{B}/m_{B}$ is the short-range
repulsive potential expressed in
terms of the bosonic scattering length $a_{B}$ ($\hbar = 1$ throughout).
Note that we have generalized the time-dependent GP equation to include a
vector potential, as a mere \emph{formal\/} device to obtain, in addition to 
the density-density correlation function already considered for neutral 
bosons, \cite{Pethick-Smith} the current-density and current-current
correlation functions. In this way we can compare these functions with the 
corresponding quantities for fermions obtained within the time-dependent
BCS approximation in the strong-coupling limit.
Accordingly, no Coulomb repulsion between bosons is  
included.

The linear-response solution of Eq.~(\ref{time-dependent-GP}) for
${\mathbf A}_{{\rm ext}}=0$ is obtained
\cite{Pethick-Smith} by taking
$
V_{{\rm ext}}({\mathbf r},t) \, = \, V_{{\mathbf q}} \, e^{i({\mathbf q}
\cdot {\mathbf r} - \omega t)} \, + \,
V_{{\mathbf q}}^{*} \, e^{-i({\mathbf q} \cdot {\mathbf r} - \omega t)}
$
with wave vector ${\mathbf q}$ and frequency $\omega$, and by setting to
linear order
\begin{equation}
\Phi({\mathbf r},t)  = \sqrt{n_{B}}\left[ 1 + {\bar u}_{\omega}({\mathbf r}) 
e^{-i \omega t}  -
 {\bar v}_{\omega}({\mathbf r})^{*}  e^{i \omega t} \right]  
e^{-i \mu_{B} t}\;.
\label{Phi-space-time}
\end{equation}
$n_{B}$ is the ground-state bosonic density, $\mu_{B} = U_{0} n_{B}$
the bosonic chemical potential, and the
``small'' components ${\bar u}_{\omega}({\mathbf r})$ and ${\bar v}_{\omega}({\mathbf r})$ are
correspondingly taken of the form
${\bar u}_{\omega}({\mathbf r}) = {\bar u}_{{\mathbf q},\omega} \, e^{i {\mathbf q} \cdot {\mathbf r}}$ and
${\bar v}_{\omega}({\mathbf r}) \, = \, {\bar v}_{{\mathbf q},\omega} \, e^{i {\mathbf q} \cdot 
{\mathbf r}}$.
The time-dependent GP equation (\ref{time-dependent-GP}) then reduces to a 
linear system for the unknowns
${\bar u}_{{\mathbf q},\omega}$ and ${\bar v}_{{\mathbf q},\omega}$.
The solution is:
\begin{equation}
{\bar u}_{{\mathbf q},\omega} = \frac{\omega  + 
E_{B}^{0}({\mathbf q})}{\omega^{2} - E_{B}({\mathbf q})^{2}} 
V_{{\mathbf q}},\; 
{\bar v}_{{\mathbf q},\omega} =  \frac{\omega - E_{B}^{0}({\mathbf q})}
{\omega^{2} - E_{B}({\mathbf q})^{2}} V_{{\mathbf q}}\; .      
\label{u-v-solution-V}
\end{equation}
$E_{B}^{0}({\mathbf q}) = {\mathbf q}^{2}/(2 m_{B})$ is the free-boson energy 
and
$E_{B}({\mathbf q}) = \sqrt{ E^{0}_{B}({\mathbf q})^{2} + 2 \mu_{B}
E^{0}_{B}({\mathbf q}) }$
the Bogoliubov dispersion relation.
The induced particle density and current are then obtained by expanding the 
expressions
$n({\mathbf r},t) = |\Phi({\mathbf r},t)|^{2}$ and
${\mathbf j}({\mathbf r},t) = {\rm Im} [ \Phi({\mathbf r},t)^{*} \nabla
\Phi({\mathbf r},t))/ m_{B}]$ to linear order in $V_{{\rm ext}}$.
One obtains:
\begin{eqnarray}
\delta n({\mathbf r},t) &=& \frac{n_{B}}{m_{B}} \,
\frac{{\mathbf q}^{2}}{\omega^{2} - E_{B}({\mathbf q})^{2}}
\,\, V_{{\rm ext}}({\mathbf r},t)
\label{n-solution-V}
\\
{\mathbf j}({\mathbf r},t) &=& \frac{n_{B}}{m_{B}} \frac{{\mathbf q} \,
\omega}{\omega^{2} - E_{B}({\mathbf q})^{2}}
 V_{{\rm ext}}({\mathbf r},t)\; .
\label{j-solution-V}
\end{eqnarray}
These satisfy the continuity equation
$\partial \delta n({\mathbf r},t)/\partial t + \nabla \cdot
{\mathbf j}({\mathbf r},t) = 0$. By linear-response theory, the
density-density and current-density correlation
functions can be identified:
\begin{eqnarray}
\chi_{n n}({\mathbf q},\omega)&=& \frac{n_{B}}{m_{B}} 
\frac{{\mathbf q}^{2}}{\omega^{2} - E_{B}({\mathbf q})^{2}}
\label{chinn}\\
\chi_{{\mathbf j} n}({\mathbf q},\omega) &=& \frac{n_{B}}{m_{B}} \,
\frac{{\mathbf q} \, \omega}{\omega^{2} - E_{B}({\mathbf q})^{2}}
\label{chijn}
\end{eqnarray}
with the typical Bogoliubov form.

Similarly, the linear-response solution of
Eq.~(\ref{time-dependent-GP}) when $V_{{\rm ext}}=0$ is obtained
by taking
${\mathbf A}_{{\rm ext}}({\mathbf r},t) \, = \, {\mathbf A}_{{\mathbf q}} \,
e^{i({\mathbf q} \cdot {\mathbf r} - \omega t)}
\, + \, {\mathbf A}_{{\mathbf q}}^{*} \, e^{-i({\mathbf q} \cdot {\mathbf r} -
\omega t)}$            
and using the same form (\ref{Phi-space-time}) for the wave function.
Working in a gauge where $\nabla \cdot
{\mathbf A}_{{\rm ext}}({\mathbf r},t)$ is nonzero (such that
${\mathbf q} \cdot {\mathbf A}_{{\mathbf q}} \ne 0)$ and retracing the above
arguments, one obtains
\begin{eqnarray}
{\bar u}_{{\mathbf q},\omega} & = & \frac{ E_{B}^{0}({{\mathbf q}})
 +  2  \mu_{B}  + \omega}
{ E_{B}({\mathbf q})^{2}  - \omega^{2}}  \frac{q_B}{2 m_{B}c} 
\, {\mathbf q} \cdot A_{{\mathbf q}}\label{u-solution}
\\
{\bar v}_{{\mathbf q},\omega} & = & \sqrt{n_{B}}\, 
\frac{ E_{B}^{0}({{\mathbf q}}) +  2 \mu_{B} -  \omega}
{E_{B}({\mathbf q})^{2}  - \omega^{2}} \frac{q_B}{2 m_{B} c}\,
 {\mathbf q} \cdot A_{{\mathbf q}}
\label{v-solution}
\end{eqnarray}
in the place of (\ref{u-v-solution-V}).
The ensuing induced density and current now take the form:
\begin{eqnarray}
&&\delta n({\mathbf r},t) = \frac{n_{B}}{m_{B}} \frac{\omega 
{\mathbf q}}{\omega^{2} - E_{B}({\mathbf q})^{2}}
\cdot \left( - \frac{q_{B}}{c}   {\mathbf A}_{{\rm ext}}({\mathbf r},t)
\right)  \label{n-solution-A}\\
&&{\mathbf j}({\mathbf r},t) = \frac{2n_{B}}{(2m_{B})^{2}} {\mathbf q} 
\frac{ E^{0}_{B}({\mathbf q})  + 2 \mu_{B}}{\omega^{2}
- E_{B}({\mathbf q})^{2}}  {\mathbf q} 
\cdot \left( - \frac{q_{B}}{c}  {\mathbf A}_{{\rm ext}}({\mathbf r},t)
\right)\; .
\label{j-solution-A}
\end{eqnarray}
These replace (\ref{n-solution-V}) and (\ref{j-solution-V}), with the 
continuity equation being
satisfied once the diamagnetic
contribution $-(q_{B} n_{B})/(m_{B} c)
{\mathbf A}_{{\rm ext}}({\mathbf r},t)$ is added to the induced
current (\ref{j-solution-A}).
The density-current and current-current correlation functions also acquire 
the typical Bogoliubov form:
\begin{eqnarray}
\chi_{n {\mathbf j}}({\mathbf q},\omega) &=& \frac{n_{B}}{m_{B}} \,
\frac{\omega \, {\mathbf q}}{\omega^{2} - E_{B}({\mathbf q})^{2}}
\label{chinj}\\
\chi_{{\mathbf j} {\mathbf j}}({\mathbf q},\omega) &=&
\frac{n_{B}}{m_{B}^{2}} \, {\mathbf q} \,
\frac{\mu_{B} \, + \, E^{0}_{B}({\mathbf q})/2}{\omega^{2} -
E_{B}({\mathbf q})^{2}} \, {\mathbf q}
\,\, .
\label{chijj}
\end{eqnarray}
The correlation functions (\ref{chinn})-(\ref{chijn}) and 
(\ref{chinj})-(\ref{chijj})
describe the linear response of the bosonic condensate to
an external disturbance in terms of Bogoliubov quasi-particle excitations.

We now show that the same form of the correlation functions is
obtained from the fermionic time-dependent BCS
(or broken-symmetry RPA) approximation in the strong-coupling limit of the
fermionic attraction, when composite bosons with
$m_{B}=2m$ and $q_{B}=2 q$ form. $m$ and $q$ are the fermionic mass and 
charge.
We first express the fermionic
correlation functions in the broken-symmetry phase
in terms of the two-particle correlation function
\begin{eqnarray}
L(1,2;1',2') &=& \langle T_{\tau} \left[ \psi(1) \psi(2) \psi^{\dagger}
(2') \psi^{\dagger} (1')\right] \rangle \nonumber\\
& - & {\mathcal G}(1,1') \, {\mathcal G}(2,2')  \,\, .
\label{L-definition}
\end{eqnarray}
Here, $\langle \cdots \rangle$ is a thermal average, $T_{\tau}$ the
time-order operator for imaginary time $\tau$,
$1=({\mathbf r}_{1},\tau_{1},\ell_{1})$ with the index $\ell=(I,II)$
identifying the components of the field operator in
the Nambu representation ($\psi_{I}({\mathbf r})=\psi_{\uparrow}({\mathbf r})$
and
$\psi_{II}({\mathbf r})=\psi_{\downarrow}^{\dagger}({\mathbf r})$),
and ${\mathcal G}(1,2) \, = \, - \, \langle T_{\tau} \left[ \psi(1)
\psi^{\dagger} (2) \right] \rangle$ is the single-particle
Green's function.
The two-particle correlation function (\ref{L-definition}) can be
expressed in terms of the
many-particle T-matrix~\cite{RNC}:
\begin{eqnarray}
& & - \, L(1,2;1',2') =  {\mathcal G}(1,2') \, {\mathcal G}(2,1')
+  \int \! d3456 \nonumber \\
& &  \times {\mathcal G}(1,3) \, {\mathcal G}(6,1') \, T(3,5;6,4)
\, {\mathcal G}(4,2') \, {\mathcal G}(2,5)\; . 
\label{L-T-relation}
\end{eqnarray}
An explicit form for $T$ results from the Bethe-Salpeter
equation once a choice for its kernel is made.

Within the (off-diagonal) BCS approximation we are
considering, the Bethe-Salpeter equation for $T$ can
be explicitly solved in the frequency and wave-vector representation.
For fermions interacting via an attractive
short-range potential regularized in terms of the
two-body scattering length $a_{F}$ (with $a_{B}=2a_{F}$)\cite{regularization},
the only nonvanishing elements of the many-particle T-matrix 
correspond to the Nambu indices reported in Fig.~1(a).
In the strong-coupling limit ($\beta \mu \rightarrow -\infty$ with inverse 
temperature $\beta$ and fermionic chemical potential 
$\mu$) of the attractive fermionic interaction,
one obtains the following expressions for the nonvanishing elements of the
many-particle T-matrix:
\begin{eqnarray}
T_{11}({\mathbf q},\Omega_{\nu}) & = & T_{22}({\mathbf q},-\Omega_{\nu}) 
\nonumber\\
&\simeq& \frac{8\pi}{m^{2}a_{F}} 
\frac{i\Omega_{\nu} +  {\mathbf q}^{2}/(2 m_{B}) +  \mu_{B}}
     {(i\Omega_{\nu})^{2}  -  E_{B}({\mathbf q})^{2}}
\label{T-11-strong-coupling} \\
T_{12}({\mathbf q},\Omega_{\nu}) & = & T_{21}({\mathbf q},\Omega_{\nu})
\nonumber\\ 
&\simeq& - \frac{8\pi}{m^{2}a_{F}} 
\frac{\mu_{B}}{(i\Omega_{\nu})^{2} \, - \,
E_{B}({\mathbf q})^{2}}\; .              \label{T-12-strong-coupling}
\end{eqnarray}
$\Omega_{\nu} = 2 \nu \pi/\beta$ ($\nu$ integer) is a bosonic
Matsubara frequency, $\mu_{B} = 2\mu + \varepsilon_{0}$
is the bosonic chemical potential (with $\varepsilon_{0} = (ma_{F}^{2})^{-1}$ 
the two-body binding energy), and $E_{B}({\mathbf q})$
coincides formally with the Bogoliubov dispersion
of Eq.~(\ref{u-v-solution-V}).
The fermionic single-particle Green's functions are correspondingly
given by the BCS expressions
\begin{eqnarray}
{\mathcal G}_{1 1}({\mathbf k},\omega_n) \, & = & - \, {\mathcal G}_{2
2}({\mathbf k},-\omega_n) \, = \,
- \frac{\xi({\mathbf k}) + i \omega_n}{E_{F}({\mathbf k})^2 + \omega_n^2}
\label{BCS-Green-function-11}  \\
{\mathcal G}_{2 1}({\mathbf k},\omega_n) \, & = & \, {\mathcal G}_{1
2}({\mathbf k},\omega_n) \, = \,
\frac{\Delta}{E_{F}({\mathbf k})^2 + \omega_n^2}\; .
\label{BCS-Green-function-12}
\end{eqnarray}
$\omega_{n} = (2n + 1) \pi /\beta$ ($n$ integer) is a fermionic
Matsubara frequency,
$\xi({\mathbf k})={\mathbf k}^{2}/(2m) - \mu$, and
$E({\mathbf k})=\sqrt{\xi({\mathbf k})^{2}+\Delta^{2}}$ 
for an ($s$-wave) isotropic gap function $\Delta$.

\begin{figure}
\centerline{\epsfig{figure=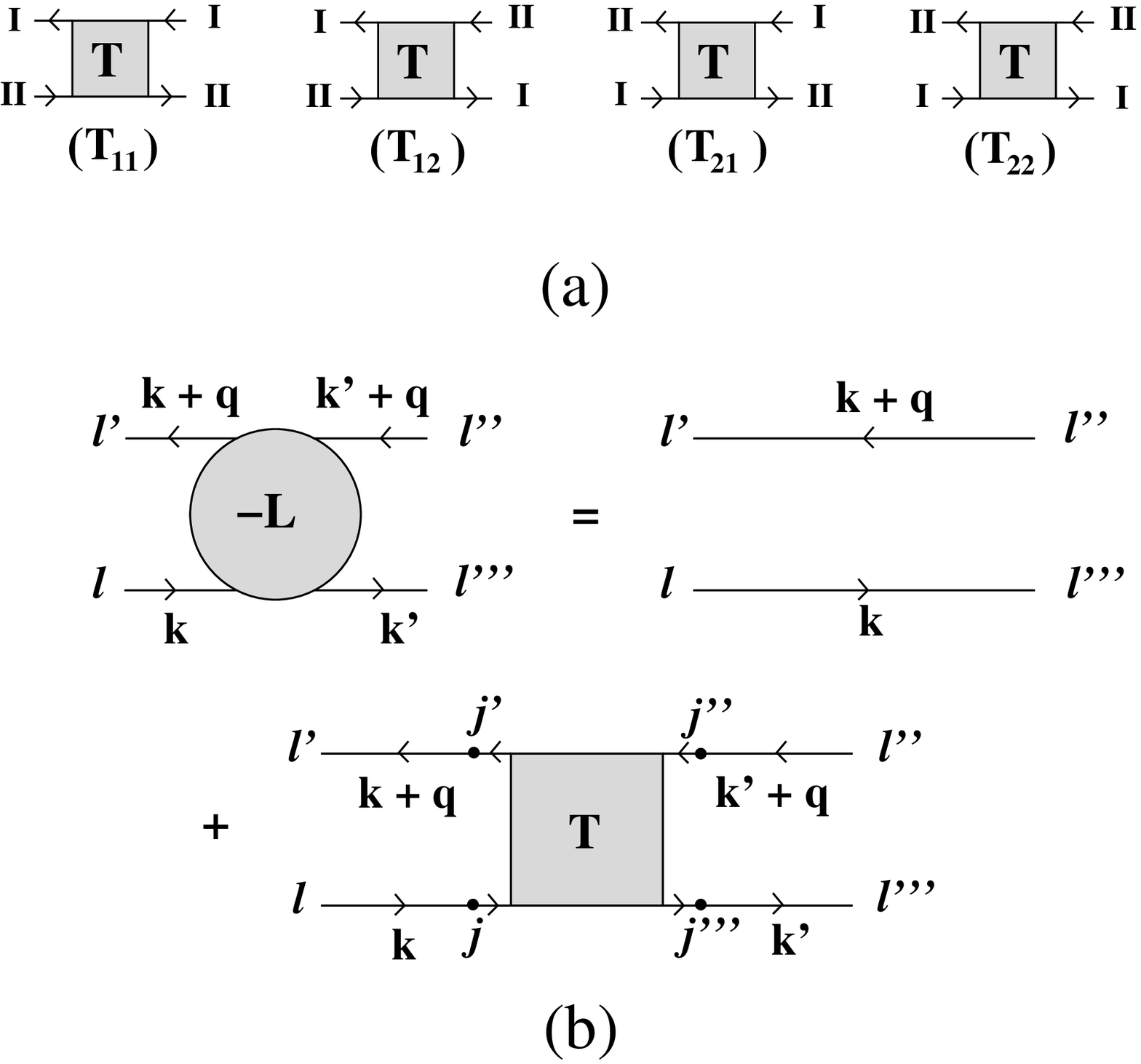,width=8.1cm}}
\vspace{.25truecm}
\caption{(a) Nambu indices associated with the four nonvanishing elements 
of the many-particle T-matrix for a fermionic contact potential;
(b) Relation between the two-particle correlation function $L$ and the
many-particle T-matrix.}
\end{figure}
The fermionic correlation functions for the number density and current can be
quite generally obtained from their definitions
as thermal averages of the time-order operator acting on the number density
and current operators. This is done by expressing the number
density and current operators in terms of Nambu field operators
and introducing the two-particle correlation function (\ref{L-definition}) 
accordingly.
One ends up with the following expressions: 
\begin{eqnarray}
& &\chi_{n n}(q) =  - \sum_{\ell,
\ell',\ell'',\ell'''}\!\tau^{3}_{\ell,\ell'} 
\tau^{3}_{\ell'',\ell'''} \int\!\frac{d{\mathbf k}\, d{\mathbf k'}}{(2\pi)^{6}} \,
\frac{1}{\beta^2} \sum_{n, n'} \nonumber\\ 
& &\times 
L^{\ell',\ell''}_{\ell,\ell'''}(k, k'; q)
\label{chi-n-n-fermions}\\
& &\chi_{n {\mathbf j}}(q) =  -  \frac{1}{2 m}
\sum_{\ell, \ell',\ell'',\ell'''}\! \tau^{3}_{\ell,\ell'} 
\tau^{0}_{\ell'',\ell'''} \int \!
\frac{d{\mathbf k} \, d{\mathbf k'}}{(2\pi)^{6}} \nonumber\\
& &\times  
(2{\mathbf k'} + {\mathbf q})\frac{1}{\beta^2} \sum_{n,n'} 
\,     
L^{\ell',\ell''}_{\ell,\ell'''}(k, k'; q)
\label{chi-n-j-fermions}\\
& &\chi_{{\mathbf j}{\mathbf j}}(q) =  - 
\frac{1}{(2 m)^{2}} \sum_{\ell, \ell',\ell'',\ell'''}\! 
\tau^{0}_{\ell,\ell'} \tau^{0}_{\ell'',\ell'''} \int \!
\frac{d{\mathbf k} \, d{\mathbf k'}}{(2\pi)^{6}} \nonumber\\
& & \times (2{\mathbf k} + {\mathbf q})(2{\mathbf k'} + {\mathbf q}) 
\frac{1}{\beta^2} \sum_{n,n'} 
L^{\ell',\ell''}_{\ell,\ell'''}(k, k'; q)\; .
\label{chi-j-j-fermions}
\end{eqnarray}
The notation is $q=({\mathbf q},\Omega_{\nu})$, 
$k=({\mathbf k},\omega_{n})$, $k'=({\mathbf k'},\omega_{n'})$,  
where $\tau^{0}$ and $\tau^{3}$ are Pauli matrices, and the labels for
$L$ are specified in Fig.~1(b).
The first term on the right-hand side of the diagrammatic
representation of Fig.~1(b) corresponds to the standard
BCS contribution to the response functions \cite{FW}. This term 
vanishes in the strong-coupling limit for temperatures
much below the dissociation temperature of the composite bosons.
The second term, on the other hand, corresponds to the
broken-symmetry BCS-RPA approximation and is required to
maintain full gauge invariance \cite{Schrieffer}.
In the present context this term has a nontrivial strong-coupling limit
and yields the response of the composite bosons condensate.
From this term one obtains for the correlation
functions (\ref{chi-n-n-fermions})-(\ref{chi-j-j-fermions}):
\begin{eqnarray}
\chi_{n n}(q) &=& \sum_{\ell,
\ell',\ell'',\ell'''} 
D_{\ell,\ell'}(q) 
T^{\ell',\ell''}_{\ell,\ell'''}(q) 
D_{\ell''',\ell''}(q)
\label{chi-n-n-approximate}\\
\chi_{n {\mathbf j}}(q) &=&\sum_{\ell,
\ell',\ell'',\ell'''} 
D_{\ell,\ell'}(q) 
T^{\ell',\ell''}_{\ell,\ell'''}(q) 
{\mathbf V}_{\ell''',\ell''}(q)
\label{chi-n-j-approximate}\\
\chi_{{\mathbf j} {\mathbf j}}(q)&=& \sum_{\ell,
\ell',\ell'',\ell'''} 
{\mathbf V}_{\ell,\ell'}(q)
T^{\ell',\ell''}_{\ell,\ell'''}(q) 
{\mathbf V}_{\ell''',\ell''}(q)
\label{chi-j-j-approximate}
\end{eqnarray}
with
\begin{eqnarray}
D_{\ell,\ell'}(q) &\equiv& \sum_{\ell'', \ell'''}
\tau^{3}_{\ell'',\ell'''} 
\int \! \frac{d{\mathbf k}}{(2\pi)^{3}} \frac{1}{\beta} \sum_{n}\nonumber\\
&\times&{\mathcal G}_{\ell,\ell''}(k) 
{\mathcal G}_{\ell''',\ell'}(q+k)
\label{D-definition}\\
{\mathbf V}_{\ell,\ell'}(q) &\equiv&
\frac{1}{2m} \, \sum_{\ell'', \ell'''} 
\tau^{0}_{\ell'',\ell'''} \int \! \frac{d{\mathbf k}}{(2\pi)^{3}} 
\frac{1}{\beta} \sum_{n}\nonumber\\ 
&\times& (2 {\mathbf k} + {\mathbf q}) 
{\mathcal G}_{\ell,\ell''}(k) \,
{\mathcal G}_{\ell''',\ell'}(q+k) .
\label{V-definition}
\end{eqnarray}
Recalling that, for a fermionic contact potential, only the
four elements (\ref{T-11-strong-coupling})
and (\ref{T-12-strong-coupling}) of the many-particle T-matrix are
nonvanishing, and exploiting the symmetry
properties of the quantities (\ref{D-definition}) and (\ref{V-definition}),
one arrives at:
\[
\chi_{n n}(q) =  \left( D(q), D(-q) \right) 
\left( \begin{array}{cc}
T_{11}(q) & T_{12}(q) \\ T_{21}(q) & T_{22}(q)
\end{array} \right)
\left( \begin{array}{c}
D(q) \\ D(-q)
\end{array} \right)
\]
\[
\chi_{n {\mathbf j}}(q) = \left(D(q), D(-q) \right)
\left( \begin{array}{cc}
T_{11}(q) & T_{12}(q) \\ T_{21}(q) & T_{22}(q)
\end{array} \right)
\left( \begin{array}{c}
{\mathbf V}(q) \\ {\mathbf V}(-q)
\end{array} \right)
\]
\[
\chi_{{\mathbf j} {\mathbf j}}(q) =  \left( {\mathbf V}(q), 
{\mathbf V}(-q) \right) 
\left( \begin{array}{cc}
T_{11}(q) & T_{12}(q) \\ T_{21}(q) & T_{22}(q)
\end{array} \right) 
\left( \begin{array}{c}
{\mathbf V}(q) \\ {\mathbf V}(-q)
\end{array} \right)
\]
where $D(q)\simeq - 4  m \Delta (m a_{F})/(16 \pi)$
and $ {\mathbf V}(q)\simeq  - {\mathbf q} \Delta (m a_{F})/(16 \pi)$.
Using the approximate expressions (\ref{T-11-strong-coupling}) and
(\ref{T-12-strong-coupling}),
one obtains finally for the fermionic correlation functions in the
strong-coupling limit:
\begin{eqnarray}
\chi_{n n}(q) & \simeq & D^{2} (q=0) \, \left[ T_{11}(q) \, + \, T_{11}(-q)
\, + \, 2 \, T_{12}(q)  \right]   \nonumber \\
& = & 4 \, \frac{n_{B}}{m_{B}} \, \frac{{\mathbf q}^{2}}{(i
\Omega_{\nu})^{2} \, - \, E_{B}({\mathbf q})^{2}}
\label{chi-n-n-final}\\
\chi_{n {\mathbf j}}(q) & \simeq & D(q=0) \, {\mathbf V}(q) \, \left[
T_{11}(q) \, - \, T_{11}(-q) \right]      \nonumber \\
& = & 4 \, \frac{n_{B}}{m_{B}} \, \frac{i \Omega_{\nu} \, {\mathbf q}}{(i
\Omega_{\nu})^{2} \, - \, E_{B}({\mathbf q})^{2}}
\label{chi-n-j-final}\\
\chi_{{\mathbf j} {\mathbf j}}(q) & \simeq & {\mathbf V}(q) \,{\mathbf V}(q) 
\, \left[ T_{11}(q) \, + \, T_{11}(-q) \, - \, 2 \, T_{12}(q) \right]
\nonumber \\
& = & 4 \, \frac{n_{B}}{m_{B}^{2}} \, {\mathbf q} \,
\frac{\mu_{B} \, + \, {\mathbf q}^{2}/(4 m_{B})}{(i
\Omega_{\nu})^{2} \, - \, E_{B}({\mathbf q})^{2}} \, {\mathbf q} \,\, .
\label{chi-j-j-final}
\end{eqnarray}
The analytic continuation $i \Omega_{\nu} \rightarrow \omega
+ i \delta$ ($\delta = 0^{+}$) to the real frequency $\omega$ is now required
to obtain the retarded correlation functions. In this way the bosonic 
expressions (\ref{chinn}),(\ref{chinj}), and (\ref{chijj}) are recovered from
their fermionic counterparts in the strong-coupling limit. 
There is an additional factor of four, reflecting the fact that the bosonic 
number and current densities are half the fermionic values.

The present mapping has been explicitly established for fermions
(and composite bosons) in the absence of a static external potential. On 
physical grounds we expect the connection we have established 
between the time-dependent GP equation and the fermionic broken-symmetry 
BCS-RPA approximation to remain valid also in the presence of a static
confining potential.

The above forms of the correlation functions obtained either from the
time-dependent GP equation or from the
strong-coupling limit of the fermionic broken-symmetry BCS-RPA
approximation, contain \emph{only\/} the contribution
of the condensate and thus hold at or near zero temperature.
Since the current response of the condensate can be only
longitudinal, the transverse current correlation
function cannot be obtained in this way since it requires inclusion of
excitations out of the condensate.
These excitations are important for determining the behavior
of the system at finite temperature
and close to the superfluid transition temperature.
Corrections to the time-dependent GP equation, or equivalently additional
fermionic contributions to the current correlation function in the 
broken-symmetry phase over and above the BCS-RPA approximation, are thus 
required to recover, for instance, the Bogoliubov approximation to the current
correlation function for bosons~\cite{Bassani-GCS}. This contains a
transverse in addition to a longitudinal part.

In conclusion, we have shown that in the strong-coupling
limit of the mutual fermionic attraction the response functions obtained at 
zero temperature from the linearized form of the
time-dependent GP equation for a system of composite bosons are equivalent to 
those obtained within the broken-symmetry BCS-RPA approximation for the 
original fermionic system.
The present mapping extends to the time-dependent case a previous result
\cite{PS-02} that showed the equivalence between
the time-independent GP equation for composite bosons and the
strong-coupling limit of the Bogoliubov-de Gennes equations
for fermions.

We thank D. Neilson for a critical reading of the manuscript.

\end{document}